\documentclass[12pt,subeqn]{article}

\usepackage{amsmath}

\renewcommand{\theequation}{\arabic{equation}}
\newcommand{\str}{\operatorname{str}}

\newcommand{\EQ}{\begin{equation}}
\newcommand{\EN}{\end{equation}}

\newcommand{\ket}[1]{\left|#1\right\rangle}      

\newcommand{\bear}{\begin{eqnarray}}
\newcommand{\ear}{\end{eqnarray}}
\newcommand{\bt} { \begin{tabular} }
\newcommand{\et}{ \end{tabular} }
\newcommand{\bc} { \begin{center} }
\newcommand{\ec}{ \end{center} }

\newcommand{\btb} { \begin{table} }
\newcommand{\etb}{ \end{table} }

\begin{document}

\topmargin 0pt
\oddsidemargin 5mm

\newpage
\setcounter{page}{0}
\begin{titlepage}
\begin{flushright}
UFSCARF-TH-12-15
\end{flushright}
\vspace{0.5cm}
\begin{center}
{\large Algebraic Bethe ansatz for an integrable $U_{q}[Sl(n|m)]$ \\ vertex model with mixed representations}\\
\vspace{1cm}
{\large G.A.P. Ribeiro and M.J. Martins} \\
\vspace{1cm}
{\em Universidade Federal de S\~ao Carlos\\
Departamento de F\'{\i}sica \\
C.P. 676, 13565-905~~S\~ao Carlos(SP), Brasil}\\
\end{center}
\vspace{0.5cm}

\begin{abstract}
We diagonalize the transfer matrix of a solvable vertex model constructed by combining the vector
representation of $U_{q}[Sl(n|m)]$ and its dual by means of the quantum inverse scattering
framework. The algebraic Bethe ansatz solution consider all
 $\frac{(n+m)!}{n!m!}$ possibilities of choosing the grading for arbitrary values of $n$ and $m$.
 This allows us to derive the transfer matrix eigenvalues and the respective Bethe ansatz equations
 for general grading choices.
\end{abstract}

\vspace{.15cm}
\centerline{PACS numbers:  05.50+q, 02.30.IK}
\vspace{.1cm}
\centerline{Keywords: Algebraic Bethe Ansatz, Lattice Models}
\vspace{.15cm}
\centerline{December 2005}

\end{titlepage}


\pagestyle{empty}

\newpage

\pagestyle{plain}
\pagenumbering{arabic}

\renewcommand{\thefootnote}{\arabic{footnote}}

\section{Introduction}

Nowadays it is well known that the concept of integrability based on the Yang-Baxter algebra \cite{KO} accommodates in its algebraic structure variables with commuting and anti-commuting rules of permutation \cite{KU1}. This framework provided us a systematic way of constructing and solving two-dimensional integrable vertex models possessing both bosonic and fermionic edge configurations. The respective Boltzmann weights are viewed as $R$-matrix solutions of the graded version of the Yang-Baxter equation based on supergroup symmetries \cite{KU2}. The associated transfer matrices can then be diagonalized algebraically by means of the quantum inverse scattering method \cite{KU2}.

An important class of the graded vertex system is that directly related to the finite dimensional representations of the $Sl(n|m)$ superalgebra \cite{KA}. The corresponding $R$-matrices have been given origin to a number of interesting one-dimensional magnets and interacting electrons systems. To name just some examples, we mention the lattice gas model with $n$ species of bosons and $m$ species of fermions \cite{LA}, the supersymmetric $t-J$ model \cite{TJ,ES}, integrable generalizations of the Hubbard model \cite{ES1,KAR} and Heisenberg spin chains \cite{KOK}.

Integrable impurities \cite{HO} can also be introduced into these systems by considering different representations of the $Sl(n|m)$ symmetry at some lattice sites. In this manner, an impurity $t-J$ model \cite{BE} and a family of doped   Heisenberg chains \cite{FRA}, related to four-dimensional representations of $Sl(2|1)$, have been proposed and solved by algebraic Bethe ansatz. One way of incorporating defects with the same Hilbert space as the bulk is by combining the vector and the dual representations of $Sl(n|m)$. This possibility has been explored not only in the context of impurity vertex models \cite{LIN} but also to model an integrable network system \cite{GA} to study the corner transfer matrix spectrum \cite{GA1}. However, as far as Bethe ansatz results are concerned, such construction has been restricted to the specific case of the rational $Sl(2|1)$ vertex model \cite{LIN}. This solution has recently been important to understand the universality class of a spin chain that alternates between the $3$ and $\bar{3}$ representations of $Sl(2|1)$ \cite{SA}.

These considerations indicate that the transfer matrix diagonalization of the vertex model whose weights involve both the vector and dual representation of the extended $U_{q}[Sl(n|m)]$ quantum superalgebra, for arbitrary $n$ and $m$, is worth to be pursued. This model contains an extra free parameter $q$ in such way that the classical $Sl(n|m)$ symmetry is recovered at the isotropic $q\rightarrow 1$ limit. Here we tackle this problem by means of the algebraic Bethe ansatz such that the many possible grading choices are considered in a unified way for arbitrary $n$ and $m$. We recall that the latter feature for general values of the number of bosons and fermions has been known to be technically involved in the literature \cite{KU2,ES1,GO}. The typical difficulty one finds is to establish a suitable formulation of the nested Bethe ansatz that takes arbitrary grading into account and therefore avoiding a case by case analysis \cite{GO}. In this paper this problem is circumvented and the transfer matrix eigenvalues as well as the corresponding Bethe ansatz equations are exhibited for any of the $\frac{(n+m)!}{n!m!}$ possible grading choices.

This paper is organized as follows. In the next section we describe the mixed vertex model whose weights alternates between the vector representation of $U_{q}[Sl(n|m)]$ and its dual. The corresponding Hamiltonians representing either impurity or alternating spin chain are explicitly derived. In section \ref{ABA} we describe the essential steps to solve the eigenvalue problem for the respective mixed transfer matrix. In section \ref{eingBA} explicit expressions for the eigenvalues and Bethe ansatz equations are presented for general grading. In Appendix A we summarize the coefficients of the Hamiltonians given in section \ref{vertex}.

\section{The vertex model}\label{vertex}

The system we are going to consider is determined by the fundamental $R$-matrix associated with the vector representation of the $U_{q}[Sl(n|m)]$ superalgebra \cite{SUP}. We recall here that this $R$-matrix is closely related to the Boltzmann weights of the so-called Perk-Schultz vertex model \cite{PS} and it is given by
\bear
\check{R}_{ab}(\lambda)&=& \sum_{\alpha=1}^{n+m} a_{\alpha}(\lambda) (-1)^{p_{\alpha}} \hat{e}_{\alpha \alpha}^{(a)} \otimes \hat{e}_{\alpha \alpha}^{(b)}
+ b(\lambda) \sum_{\stackrel{\alpha,\beta=1}{\alpha \neq \beta}}^{n+m} (-1)^{p_{\alpha} p_{\beta}} \hat{e}_{\alpha \beta}^{(a)} \otimes \hat{e}_{\beta \alpha}^{(b)} \nonumber \\
&+& c_{1}(\lambda) \sum_{\stackrel{\alpha,\beta=1}{\alpha < \beta}}^{n+m} \hat{e}_{\alpha \alpha}^{(a)} \otimes \hat{e}_{\beta \beta}^{(b)} + c_{2}(\lambda) \sum_{\stackrel{\alpha,\beta=1}{\alpha > \beta}}^{n+m} \hat{e}_{\alpha \alpha}^{(a)} \otimes \hat{e}_{\beta \beta}^{(b)},
\label{rmatr}
\ear
where the Boltzmann weights $a_{\alpha}(\lambda)$, $b(\lambda)$ and $c_{i}(\lambda)$ are determined by
\begin{subequations}
\bear
a_{\alpha}(\lambda)&=& \frac{q^{2 (1-p_{\alpha})}-q^{2 p_{\alpha}} e^{2\lambda} }{q^2-e^{2\lambda}}, ~~~ \alpha=1,\dots, n+m, \label{boltzWa}\\
b(\lambda)&=&\frac{q(1-e^{2\lambda})}{q^2-e^{2\lambda}}, \label{boltzWb}\\
c_{i}(\lambda)&=&\frac{(q^2-1)e^{2\lambda(2-i)}}{q^2-e^{2\lambda}} , ~~~ i=1,2 \label{boltzWc}
\ear
\end{subequations}
where $\hat{e}_{\alpha \beta}^{(a)} \in C_{a}^{n+m}$ are the standard $(n+m)\times (n+m)$ Weyl matrices. The symbols $q$ and $\lambda$ denote the quantum group deformation parameter and the spectral parameter, respectively. The Grassmann parities $p_{\alpha}$ distinguish the bosonic $p_{\alpha}=0$ and fermionic $p_{\alpha}=1$ degrees of freedom. We stress that in this paper the grading ordering of the many possible permutations of $p_{1} \dots p_{n+m}$  will be considered arbitrary.

The graded Yang-Baxter \cite{KU1} satisfied by the $R$-matrix (\ref{rmatr}) is the basic object to build the transfer matrices of solvable vertex models mixing between different representations of $U_{q}[Sl(n|m)]$. This is an associative algebra generated by the elements of the Lax operators ${\cal L}_{{\cal A}j}(\lambda)$, namely
\EQ
\check{R}(\lambda-\mu) {\cal L}_{{\cal A}j}(\lambda) \stackrel{s}{\otimes} {\cal L}_{{\cal A}j}(\mu) =  {\cal L}_{{\cal A}j}(\mu) \stackrel{s}{\otimes} {\cal L}_{{\cal A}j}(\lambda) \check{R}(\lambda-\mu),
\label{alg}
\EN
where the supertensor product $A\stackrel{s}{\otimes} B$ between two matrices with elements $A_{\alpha\beta}$ and $B_{\gamma\delta}$ is viewed as $\displaystyle (A \stackrel{s}{\otimes} B)_{12}=\sum_{\alpha\beta\gamma\delta=1}^{n+m} (-1)^{p_{\beta} (p_{\alpha}+p_{\gamma})} A_{\alpha \gamma} B_{\beta \delta} \hat{e}_{\alpha \gamma}^{(1)}\otimes \hat{e}_{\beta \delta}^{(2)}$.

The operators ${\cal L}_{{\cal A}j}(\lambda)$ are matrices on the auxiliary space ${\cal A}\equiv {\cal C}^{n+m}$  whose elements are operators on the quantum space $V_{j} \in U_q[Sl(n|m)]$. The simplest Lax operator occurs when $V_{j}$ is chosen to be the space of the vector representation of $U_{q}[Sl(n|m)]$. In this case the algebra (\ref{alg}) coincides with the Yang-Baxter equation for $\check{R}_{ab}(\lambda)$ provided we set
\EQ
{\cal L}_{ab}^{(1)}(\lambda)=P_{ab} \check{R}_{ab}(\lambda),
\EN
where $P_{ab}$ is the graded permutation operator $\displaystyle P_{ab}=\sum_{\alpha, \beta=1}^{n+m} (-1)^{p_{\alpha}p_{\beta}} \hat{e}_{\alpha \beta}^{(a)} \otimes \hat{e}_{\beta \alpha}^{(b)}$.

The other possibility we are interested in is the situation in which $V_{j}$ is taken as the dual of the vector representation. The general theory concerning the construction of universal Lax operators for the $U_{q}[Sl(n|m)]$ superalgebra have already been discussed in ref. \cite{ZA}. To obtain from that suitable expressions for a posteriori Bethe ansatz analysis requires, however, a certain amount of additional work. Omitting here these steps we find that the Lax operator intertwining between the vector and conjugated representation of $U_{q}[Sl(n|m)]$ in the Weyl basis is
\bear
{\cal L}_{ab}^{(2)}(\lambda) &=& \sum_{\alpha=1}^{n+m} \bar{a}_{\alpha}(\lambda) \hat{e}_{\alpha \alpha}^{(a)} \otimes \hat{e}_{\alpha \alpha}^{(b)}
+ \bar{b}(\lambda) \sum_{\stackrel{\alpha,\beta=1}{\alpha \neq \beta}}^{n+m} \hat{e}_{\alpha \alpha}^{(a)} \otimes \hat{e}_{\beta \beta}^{(b)} \nonumber \\
&+& \bar{c}_{1}(\lambda) \sum_{\stackrel{\alpha,\beta=1}{\alpha < \beta}}^{n+m} (-1)^{p_{\alpha}} q^{-2 \delta_{\alpha,1}} \hat{e}_{\alpha \beta}^{(a)} \otimes \hat{e}_{\alpha \beta}^{(b)}
+ \bar{c}_{2}(\lambda) \sum_{\stackrel{\alpha,\beta=1}{\alpha > \beta}}^{n+m} (-1)^{p_{\alpha}} q^{2 (\delta_{\beta,1}-1)} \hat{e}_{\alpha \beta}^{(a)} \otimes \hat{e}_{\alpha \beta}^{(b)},
\ear
where the new Boltzmann weights $\bar{a}_{\alpha}(\lambda)$, $\bar{b}(\lambda)$ and $\bar{c}_{i}(\lambda)$ are given by
\begin{subequations}
\bear
\bar{a}_{\alpha}(\lambda)&=& \frac{q^{2 p_{\alpha}}-q^{-2 p_{\alpha}}e^{2\lambda}}{1-e^{2\lambda}}, ~~~ \alpha=1,\dots, n+m, \\
\bar{b}(\lambda)&=&\frac{1}{q}\frac{(q^{2}-e^{2\lambda})}{1-e^{2\lambda}},\\
\bar{c}_{i}(\lambda)&=& \frac{(1-q^2)}{1-e^{2\lambda}} e^{2\lambda(i-1)}, ~~~ i=1,2.
\ear
\end{subequations}

Having found two different Lax operators satisfying the Yang-Baxter algebra with the same $R$-matrix, an integrable vertex model that combines $L_{1}$ operators of type ${\cal L}_{{\cal A}j}^{(1)}(\lambda)$ and $L_{2}$ operators of type ${\cal L}_{{\cal A}j}^{(2)}(\lambda)$ can be directly build up by means of the quantum inverse scattering framework. As usual the corresponding row-to-row transfer matrix $T^{(L_{1},L_{2})}(\lambda)$ is written as the supertrace over the auxiliary space of an ordered operator called monodromy matrix ${\cal T}_{\cal A}^{(L_{1},L_{2})}(\lambda)$, namely
\EQ
T^{(L_{1},L_{2})}(\lambda)=\str_{\cal A}{\left[ {\cal T}_{\cal A}^{(L_{1},L_{2})}(\lambda) \right]}=\sum_{i=1}^{n+m}(-1)^{p_{i}} \left[{{\cal T}_{\cal A}^{(L_{1},L_{2})}(\lambda)}\right]_{ii},
\label{transf}
\EN
where the monodromy matrix is given by
\EQ
{\cal T}_{\cal A}^{(L_{1},L_{2})}(\lambda)=  \overline{\cal L}_{{\cal A} L_{1}+L_{2}}(\lambda) \overline{\cal L}_{{\cal A} L_{1}+L_{2}-1}(\lambda) \dots \overline{\cal L}_{{\cal A}2}(\lambda) \overline{\cal L}_{{\cal A}1}(\lambda).
\label{monodromyover}
\EN

The new operator $\overline{{\cal L}}_{{\cal A}j}(\lambda)$ is now defined as follow
\EQ
\overline{\cal L}_{{\cal A}j}(\lambda)= \begin{cases}
{\cal L}_{{\cal A}j}^{(1)}(\lambda), & \text{if } j \in \{\xi_{1}, \dots, \xi_{L_{1}}\} \\
{\cal L}_{{\cal A}j}^{(2)}(\lambda+\rho), & \text{if } j \notin \{\xi_{1}, \dots, \xi_{L_{1}}\},
\end{cases}
\label{part}
\EN
where the partition $\{ \xi_{1}\dots \xi_{L_{1}} \}$ denotes a set of integers indices assuming values on the interval $1 \leq \xi_{j} \leq L=L_{1}+L_{2}$. The inhomogeneity $\rho$ can be arbitrarily chosen and it does not spoil basic properties such as the symmetry and locality of the interaction.

Next we turn our attention to interesting types of quantum spin chains that can be derived from the transfer matrix (\ref{transf}). Though the integrability does not depend on how we choose the partition $\{ \xi_{1}\dots \xi_{L_{1}} \}$ the construction of local conserved charges commuting with $T^{(L_{1},L_{2})}(\lambda)$ does. One interesting situation occurs when we have only one operator of type ${\cal L}_{{\cal A}L}^{(2)}(\lambda)$ sitting at the end of the chain. This plays the role of an impurity on a chain of $L-1$ ${\cal L}_{{\cal A}j}^{(1)}(\lambda)$ Boltzmann weights \cite{HO}. The respective impurity Hamiltonian ${\cal H}^{(1)}$ is obtained as the logarithmic derivative of the transfer matrix (\ref{transf}) at the regular point $\lambda=0$, and reads
\bear
{\cal H}^{(1)}&=&\sum_{i=1}^{L-2} \left[{\cal L}_{i,i+1}^{(1)}(0) \right]^{-1} {\cal \dot{L}}_{i,i+1}^{(1)}(0) + \left[{\cal L}_{L-1,L}^{(2)}(\rho)\right]^{-1} {\cal \dot{L}}_{L-1,L}^{(2)}(\rho) \nonumber \\
&+& \left[{\cal L}_{L-1,L}^{(2)}(\rho) \right]^{-1} \left[ {\cal L}_{L-1,1}^{(1)}(0) \right]^{-1} {\cal \dot{L}}_{L-1,1}^{(1)}(0) {\cal L}_{L-1,L}^{(2)}(\rho).
\label{h1}
\ear

To obtain explicit expressions for this operator one needs to known the form of the inverse of ${\cal L}_{{\cal A}j}^{(2)}(\lambda)$. Since this operator combines different representations, its inverse does not follows directly from unitarity. After some cumbersome algebraic manipulations we obtain that $\left[{\cal L}_{{\cal A}j}^{(2)}(\lambda)\right]^{-1}$ is given by
\bear
\left[{\cal L}_{ab}^{(2)}(\lambda)\right]^{-1}&=& \sum_{\alpha=1}^{n+m} \underbar{$a$}_{\alpha}(\lambda) \hat{e}_{\alpha \alpha}^{(a)} \otimes \hat{e}_{\alpha \alpha}^{(b)}
+ \underbar{$b$}(\lambda) \sum_{\stackrel{\alpha,\beta=1}{\alpha \neq \beta}}^{n+m}  \hat{e}_{\alpha \alpha}^{(a)} \otimes \hat{e}_{\beta \beta}^{(b)}  \label{inver} \\
&+& \underbar{$c$}_{1}(\lambda) \sum_{\stackrel{\alpha,\beta=1}{\alpha < \beta}}^{n+m}  (-1)^{p_{\alpha}} \zeta_{\alpha,\beta} \hat{e}_{\alpha \beta}^{(a)} \otimes \hat{e}_{\alpha \beta}^{(b)}
+ \underbar{$c$}_{2}(\lambda) \sum_{\stackrel{\alpha,\beta=1}{\alpha > \beta}}^{n+m}  \frac{(-1)^{p_{\alpha}}}{\zeta_{\beta,\alpha}}  \hat{e}_{\alpha \beta}^{(a)} \otimes \hat{e}_{\alpha \beta}^{(b)} , \nonumber
\ear
where the functions $\underbar{$a$}_{\alpha}(\lambda)$, $\underbar{$b$}(\lambda)$ and $\underbar{$c$}_{i}(\lambda)$ are
\begin{subequations}
\bear
\underbar{$a$}_{\alpha}(\lambda)&=&  q^{2(1-p_{\alpha})} \frac{(1-e^{2\lambda})(1-e^{2\lambda} q^{2(n-m-2+2 p_{\alpha})})}{(q^2-e^{2\lambda})(1-e^{2\lambda} q^{2(n-m-1)})}, ~~~ \alpha=1,\dots, n+m, \\
\underbar{$b$}(\lambda)&=& \frac{q(1-e^{2\lambda})}{ (q^{2}-e^{2\lambda})}  ,\\
\underbar{$c$}_{i}(\lambda)&=& \frac{(q^2-1)(1-e^{2\lambda})q^{2(n-m-1)}}{ (q^{2}-e^{2\lambda})(1-e^{2\lambda} q^{2(n-m-1)})}
e^{2\lambda(i-1)} , ~~~ i=1,2,
\ear
\end{subequations}
and
\EQ
\zeta_{i,j}=q^{-2(i-j+\delta_{i,1})} q^{2(p_{i}+p_{j})} \prod_{k=i}^{j} q^{-4 p_{k}},
\EN
for $i<j$, $i,j=1,\dots,n+m$.

Substituting Eq.(\ref{inver}) into Eq.(\ref{h1}) and after several simplifications we find that ${\cal H}^{(1)}$ can be rewritten as
\EQ
{\cal H}^{(1)}=\sum_{i=1}^{L-2} h^{(1)}_{i,i+1}  + h^{(2)}_{L-1,L} + h^{(3)}_{L-1,L,1},
\EN
where $h^{(1)}_{i,i+1}$, $h^{(2)}_{i,i+1}$ and $h^{(3)}_{L,L-1,1}$ are defined as follows
\bear
h^{(1)}_{i,i+1}&=&\dot{\check{R}}_{i,i+1}(0) ,\\
h^{(2)}_{i,i+1}&=&\sum_{\alpha=1}^{n+m} \underbar{$a$}_{\alpha}(\rho)\dot{\bar{a}}_{\alpha}(\rho) \hat{e}_{\alpha \alpha}^{(i)} \otimes \hat{e}_{\alpha \alpha}^{(i+1)} + \sum_{\stackrel{\alpha,\beta=1}{\alpha \neq \beta}}^{n+m} \underbar{$b$}(\rho)\dot{\bar{b}}(\rho) \hat{e}_{\alpha \alpha}^{(i)} \otimes \hat{e}_{\beta \beta}^{(i+1)} + \sum_{\alpha,\beta=1}^{n+m} \Theta_{\alpha,\beta} \hat{e}_{\alpha \beta}^{(i)} \otimes \hat{e}_{\alpha \beta}^{(i+1)},
\ear
\bear
h^{(3)}_{L,L-1,1}&=&\sum_{\alpha=1}^{n+m}\underbar{$a$}_{\alpha}(\rho)\dot{a}_{\alpha}(0)\bar{a}_{\alpha}(\rho)(-1)^{p_{\alpha}} \hat{e}_{\alpha \alpha}^{(L-1)} \otimes \hat{e}_{\alpha \alpha}^{(L)} \otimes \hat{e}_{\alpha \alpha}^{(1)}  \nonumber \\
&+& \sum_{\alpha,\beta=1}^{n+m} \underbar{$b$}(\rho)\dot{a}_{\alpha}(0)\bar{b}(\rho) (-1)^{p_{\alpha}} \hat{e}_{\alpha \alpha}^{(L-1)} \otimes \hat{e}_{\beta \beta}^{(L)} \otimes \hat{e}_{\alpha \alpha}^{(1)} \nonumber \\
&+&\sum_{\alpha,\beta=1}^{n+m} \Omega^{(1)}_{\alpha,\beta} \hat{e}_{\alpha \alpha}^{(L-1)} \otimes \hat{e}_{\alpha\alpha}^{(L)} \otimes \hat{e}_{\beta\beta}^{(1)} + \sum_{\alpha,\beta=1}^{n+m} \Omega^{(2)}_{\alpha,\beta} \hat{e}_{\alpha \beta}^{(L-1)} \otimes \hat{e}_{\alpha\beta}^{(L)} \otimes \hat{e}_{\alpha\alpha}^{(1)} \nonumber \\ &+&\sum_{\alpha,\beta=1}^{n+m} \Omega^{(3)}_{\alpha,\beta} \hat{e}_{\alpha \beta}^{(L-1)} \otimes \hat{e}_{\alpha\beta}^{(L)} \otimes \hat{e}_{\beta\beta}^{(1)} + \sum_{\alpha,\beta,\gamma=1}^{n+m} \Xi_{\alpha,\beta,\gamma} \hat{e}_{\alpha \beta}^{(L-1)} \otimes \hat{e}_{\alpha\beta}^{(L)} \otimes \hat{e}_{\gamma \gamma}^{(1)} \\
&+& \sum_{\alpha,\beta,\gamma=1}^{n+m} \Gamma^{(1)}_{\alpha,\beta,\gamma} \hat{e}_{\alpha \alpha}^{(L-1)} \otimes \hat{e}_{\gamma\gamma}^{(L)} \otimes \hat{e}_{\beta \beta}^{(1)} + \sum_{\alpha,\beta,\gamma=1}^{n+m} \Gamma^{(2)}_{\alpha,\beta,\gamma} M_{\alpha,\beta} . \hat{e}_{\alpha \beta}^{(L-1)} \otimes \hat{e}_{\gamma\gamma}^{(L)} \otimes \hat{e}_{\beta \alpha}^{(1)}  \nonumber\\
&+& \sum_{\alpha,\beta=1}^{n+m} \Gamma^{(3)}_{\alpha,\beta} M_{\alpha,\beta}.\hat{e}_{\alpha \beta}^{(L-1)} \otimes \hat{e}_{\gamma\beta}^{(L)} \otimes \hat{e}_{\gamma\alpha}^{(1)}  + \sum_{\alpha,\beta,\gamma=1}^{n+m} \Gamma^{(4)}_{\alpha,\beta,\gamma} M_{\alpha,\beta}.\hat{e}_{\alpha \beta}^{(L-1)} \otimes \hat{e}_{\alpha\gamma}^{(L)} \otimes \hat{e}_{\beta \gamma}^{(1)}, \nonumber
\ear
such that the coefficients $\Omega^{(i)}_{\alpha,\beta}$, $\Gamma^{(i)}_{\alpha,\beta,\gamma}$ and $\Xi_{\alpha,\beta,\gamma}$ are exhibited in Appendix A. The diagonal matrix $M_{\alpha\beta}$ has the following form
\EQ
M_{\alpha,\beta}=\sum_{\nu_{1},\dots,\nu_{L-3}=1}^{n+m} (-1)^{(p_{\alpha}+p_{\beta})\displaystyle\sum_{k=1}^{L-3}p_{\nu_{k}}} \hat{e}^{(2)}_{\nu_{1}\nu_{1}} \otimes \hat{e}^{(3)}_{\nu_{2}\nu_{2}} \otimes \dots \otimes \hat{e}^{(L-2)}_{\nu_{L-3}\nu_{L-3}}.
\EN

Yet another relevant case occurs when the number of operators ${\cal L}_{{\cal A}j}^{(1)}(\lambda)$ and ${\cal L}_{{\cal A}j}^{(2)}(\lambda)$ are equally distributed $L_{1}=L_{2}=L/2$ in an alternating way in the monodromy matrix \cite{DEG,GA}. In this situation the respective Hamiltonian is a bit more involved and is given by
\bear
{\cal H}^{(2)}&=&\sum_{\stackrel{i=1}{\text{odd } i}}^{L-1} \left[{\cal L}_{i,i+1}^{(2)}(\rho) \right]^{-1} {\cal \dot{L}}_{i,i+1}^{(2)}(\rho) + \sum_{\stackrel{i=2}{\text{even } i}}^{L} \left[{\cal L}_{i-1,i}^{(2)}(\rho) \right]^{-1} \left[ {\cal L}_{i-1,i+1}^{(1)}(0) \right]^{-1} {\cal \dot{L}}_{i-1,i+1}^{(1)}(0)  {\cal L}_{i-1,i}^{(2)}(\rho). \nonumber \\
\ear

Its explicit expression in terms of the Weyl basis, after some algebraic simplifications, involves many of the terms entering in the previous Hamiltonian. More precisely, the final result is
\EQ
{\cal H}^{(2)}=\sum_{\stackrel{i=1}{\text{odd } i}}^{L-1} h^{(2)}_{i,i+1} + \sum_{\stackrel{i=2}{\text{even } i}}^{L-2} 
\bar{h}^{(3)}_{i-1,i,i+1}  +  h^{(3)}_{L-1,L,1},
\EN
where the new bulk term $\bar{h}^{(3)}_{i-1,i,i+1}$ is given by
\bear
\bar{h}^{(3)}_{i-1,i,i+1}&=&\sum_{\alpha=1}^{n+m} \underbar{$a$}_{\alpha}(\rho)\dot{a}_{\alpha}(0)\bar{a}_{\alpha}(\rho)(-1)^{p_{\alpha}} \hat{e}_{\alpha \alpha}^{(i-1)} \otimes \hat{e}_{\alpha \alpha}^{(i)} \otimes \hat{e}_{\alpha \alpha}^{(i+1)} \nonumber \\
&+&\sum_{\alpha,\beta=1}^{n+m} \underbar{$b$}(\rho)\dot{a}_{\alpha}(0)\bar{b}(\rho) (-1)^{p_{\alpha}} \hat{e}_{\alpha \alpha}^{(i-1)} \otimes \hat{e}_{\beta \beta}^{(i)} \otimes \hat{e}_{\alpha \alpha}^{(i+1)} \nonumber \\
&+&\sum_{\alpha,\beta=1}^{n+m} \Omega^{(1)}_{\alpha,\beta} \hat{e}_{\alpha \alpha}^{(i-1)} \otimes \hat{e}_{\alpha\alpha}^{(i)} \otimes \hat{e}_{\beta\beta}^{(i+1)} + \sum_{\alpha,\beta=1}^{n+m} \Omega^{(2)}_{\alpha,\beta} \hat{e}_{\alpha \beta}^{(i-1)} \otimes \hat{e}_{\alpha\beta}^{(i)} \otimes \hat{e}_{\alpha\alpha}^{(i+1)} \nonumber \\ &+&\sum_{\alpha,\beta=1}^{n+m} \Omega^{(3)}_{\alpha,\beta} \hat{e}_{\alpha \beta}^{(i-1)} \otimes \hat{e}_{\alpha\beta}^{(i)} \otimes \hat{e}_{\beta\beta}^{(i+1)} + \sum_{\alpha,\beta,\gamma=1}^{n+m} \Xi_{\alpha,\beta,\gamma} \hat{e}_{\alpha \beta}^{(i-1)} \otimes \hat{e}_{\alpha\beta}^{(i)} \otimes \hat{e}_{\gamma \gamma}^{(i+1)} \\
&+& \sum_{\alpha,\beta,\gamma=1}^{n+m} \Gamma^{(1)}_{\alpha,\beta,\gamma} \hat{e}_{\alpha \alpha}^{(i-1)} \otimes \hat{e}_{\gamma\gamma}^{(i)} \otimes \hat{e}_{\beta \beta}^{(i+1)} + \sum_{\alpha,\beta,\gamma=1}^{n+m} \Gamma^{(2)}_{\alpha,\beta,\gamma} (-1)^{(p_{\alpha}+p_{\beta})p_{\gamma}} \hat{e}_{\alpha \beta}^{(i-1)} \otimes \hat{e}_{\gamma\gamma}^{(i)} \otimes \hat{e}_{\beta \alpha}^{(i+1)}  \nonumber\\
&+& \sum_{\alpha,\beta=1}^{n+m} \Gamma^{(3)}_{\alpha,\beta} (-1)^{(p_{\alpha}+p_{\gamma})p_{\gamma}}\hat{e}_{\alpha \beta}^{(i-1)} \otimes \hat{e}_{\gamma\beta}^{(i)} \otimes \hat{e}_{\gamma\alpha}^{(i+1)}  \nonumber \\
&+& \sum_{\alpha,\beta,\gamma=1}^{n+m} \Gamma^{(4)}_{\alpha,\beta,\gamma} (-1)^{(p_{\gamma}+p_{\beta})p_{\gamma}}\hat{e}_{\alpha \beta}^{(i-1)} \otimes \hat{e}_{\alpha\gamma}^{(i)} \otimes \hat{e}_{\beta \gamma}^{(i+1)}. \nonumber
\ear

From the form of $\left[{\cal L}_{ab}^{(2)}(\lambda+\rho)\right]^{-1}$ one clearly sees that the above expressions for ${\cal H}^{(1)}$ and ${\cal H}^{(2)}$ for $\rho=0$ make sense only when $n-m\neq 1$. In fact, for generic values of $q$ we observe $\det{\left[(1-e^{2\lambda}){\cal L}_{ab}^{(2)}(\lambda) \right]}|_{\lambda=0}$ vanishes for $n-m=1$. It should be remarked, however, that such singularity is easily avoided in the rational limit  $q\rightarrow 1$.

\section{Algebraic Bethe ansatz}\label{ABA}

The purpose of this section is to describe the essential tools entering in the diagonalization problem of the transfer matrix (\ref{transf}), namely
\EQ
T^{(L_{1},L_{2})}(\lambda) \ket{\phi} = \Lambda(\lambda) \ket{\phi},
\label{aut}
\EN
by means of the nested Bethe ansatz approach \cite{KU4}.

An important requirement in this method is the existence of an appropriate reference state $\ket{0}$ with the following properties. The action of the monodromy operator on this state gives us a triangular matrix as well as it is an exact eigenvector of the diagonal elements of ${\cal T}_{{\cal A}}^{(L_{1},L_{2})}(\lambda)$. The structure of the $\cal L$-operators of previous section suggests that such state can be built up by the tensor product
\EQ
\ket{0}=\prod_{j \in \{\xi_{1},\dots,\xi_{L_{1}} \}} \otimes \ket{0^{(1)}}_{j} \otimes \prod_{j\notin \{\xi_{1},\dots,\xi_{L_{1}} \}} \otimes \ket{0^{(2)}}_{j},
\EN
where the local pseudovacuums $\ket{0^{(i)}}_{j}$ $(i=1,2)$ acting on the $j$-th site of the $L_{1}+L_{2}$ chain are the following vectors with length $(n+m)$
\EQ
\ket{0^{(1)}}=\left( \begin{array}{c}
1 \\
0 \\
\vdots
\\
0
\end{array} \right)_{(n+m)}, \ket{0^{(2)}}=\left( \begin{array}{c}
0 \\
\vdots \\
0 \\
1
\end{array} \right)_{(n+m)}.
\EN

It is not difficult to see that the action of each operator ${\cal L}_{{\cal A}j}^{(i)}(\lambda)$ on its respective local state $\ket{0^{(i)}}_{j}$ gives as result an upper triangular matrix, namely
\EQ
{\cal L}_{{\cal A}j}^{(1)}(\lambda) \ket{0^{(1)}}_{j}={
\left( \begin{array}{ccccc}
f_{1}(\lambda) & \# & \cdots & \# & \# \\
0 & f_{2}(\lambda) & \cdots & 0 & 0 \\
\vdots & \vdots & \ddots & \vdots & \vdots \\
0 & 0 & \cdots & f_{n+m-1}(\lambda) & 0 \\
0 & 0 & \cdots & 0 & f_{n+m}(\lambda)
\end{array}\right) \ket{0^{(1)}}_{j}}_{(n+m)\times (n+m)},
\label{triang1}
\EN
and
\EQ
{\cal L}_{{\cal A}j}^{(2)}(\lambda+\rho) \ket{0^{(2)}}_{j}={
\left( \begin{array}{ccccc}
g_{1}(\lambda) & 0 & \cdots & 0 & \# \\
0 & g_{2}(\lambda) & \cdots & 0 & \# \\
\vdots & \vdots & \ddots & \vdots & \vdots \\
0 & 0 & \cdots & g_{n+m-1}(\lambda) & \# \\
0 & 0 & \cdots & 0 & g_{n+m}(\lambda)
\end{array}\right) \ket{0^{(2)}}_{j}}_{(n+m)\times (n+m)},
\label{triang2}
\EN
where the symbol $\#$ stands for non-null values. The functions $f_{i}(\lambda)$ and  $g_{i}(\lambda)$ are given by
\EQ
f_{i}(\lambda)=
\begin{cases}
a_{1}(\lambda), & i=1, \\
b(\lambda), & i=2,\dots,n+m
\end{cases}, ~~~
g_{i}(\lambda)=
\begin{cases}
\bar{b}(\lambda+\rho), & i=1,\dots,n+m-1 \\
\bar{a}_{n+m}(\lambda+\rho), & i=n+m.
\end{cases}
\EN

The next step is to write a suitable ansatz for the monodromy matrix ${\cal T}_{\cal A}^{(L_{1},L_{2})}(\lambda)$ in the auxiliary space $\cal A$. The above triangular properties suggest us to seek for the structure used in the nested Bethe ansatz diagonalization of $SU(n)$ vertex models \cite{KU4},
\EQ
{\cal T}_{\cal A}(\lambda)=\left(
\begin{array}{cccc}
A(\lambda) & B_{1}(\lambda) & \cdots & B_{n+m-1}(\lambda) \\
C_{1}(\lambda) & D_{1,1}(\lambda) & \cdots & D_{1, n+m-1}(\lambda) \\
\vdots & \vdots & \ddots & \cdots \\
C_{n+m-1}(\lambda) & D_{n+m-1, 1}(\lambda) & \cdots & D_{n+m-1, n+m-1}(\lambda)
\end{array}\right)_{(n+m)\times (n+m)},
\EN
and in terms of this representation the eigenvalue problem (\ref{aut}) becomes
\EQ
\left[(-1)^{p_{1}} A(\lambda) + \sum_{i=1}^{n+m-1} (-1)^{p_{i+1}} D_{ii}(\lambda) \right] \ket{\phi} = \Lambda(\lambda) \ket{\phi}.
\label{aut2}
\EN

An immediate consequence of properties (\ref{triang1}) and (\ref{triang2}) is that one can easily derive how the elements of the monodromy matrix acts on the reference state $\ket{0}$. The action of the diagonal fields turns out to be
\bear
A(\lambda) \ket{0} &=& \left[f_{1}(\lambda) \right]^{L_{1}} \left[ g_{1}(\lambda) \right]^{L_{2}} \ket{0}, \\
D_{ii}(\lambda) \ket{0} &=& \left[ f_{i+1}(\lambda) \right]^{L_{1}} \left[ g_{i+1}(\lambda) \right]^{L_{2}} \ket{0}, ~~~  i=1,\dots,n+m-1,
\ear
while that of the non-diagonal operators are
\bear
&&B_{i}(\lambda) \ket{0} = \#, ~~~~ D_{i,j}(\lambda) \ket{0} = 0, ~~~ i\neq j, j \neq n+m-1, \\
&&C_{i}(\lambda) \ket{0} = 0, ~~~~ D_{i,n+m-1}(\lambda) \ket{0} = \# , ~~~ i \neq n+m-1.
\ear

This reveals us that $B_{i}(\lambda)$ are creation fields with respect to the reference state $\ket{0}$. The next step in the algebraic Bethe ansatz method is to search for other transfer matrix eigenvectors as linear combinations of products of such creation fields acting on $\ket{0}$, namely
\EQ
\ket{\phi}=B_{\alpha_{1}}(\lambda_{1}^{(1)}) \dots B_{\alpha_{\bar{m}_{1}}}(\lambda_{\bar{m}_{1}}^{(1)}) {\cal F}^{\alpha_{\bar{m}_{1}}\dots \alpha_{1}} \ket{0},
\label{ansatz}
\EN
where sum over repeated indices $\alpha_{k}=1,\dots, n+m-1$ is assumed and ${\cal F}^{\alpha_{\bar{m}_{1}}\dots \alpha_{1}}$ are coefficients of the linear combination.

In order to solve the problem (\ref{aut2}) one has to carry on the fields $A(\lambda)$ and $D_{ii}(\lambda)$ over the operators $B_{i}(\lambda_{i}^{(1)})$ until they reach the reference state $\ket{0}$. This is done with the help of an appropriate set of commutation rules coming from the Yang-Baxter (\ref{alg}) for ${\cal T}_{\cal A}^{(L_{1},L_{2})}(\lambda)$. The most useful relations are
\begin{subequations}
\bear
A(\lambda) B_{j}(\mu) &=& \frac{a_{1}(\mu-\lambda)}{b(\mu-\lambda)} B_{j}(\mu) A(\lambda)  -\frac{c_{2}(\mu-\lambda)}{b(\mu-\lambda)} B_{j}(\lambda) A(\mu) (-1)^{p_{1}}, \label{rule1}\\
D_{ij}(\lambda) B_{k}(\mu) &=& \frac{1}{b(\lambda-\mu)} B_{l}(\mu) D_{iq}(\lambda) \check{r}^{(1)}(\lambda-\mu)_{lq}^{jk} (-1)^{p_{i+1}p_{l+1}+p_{1}(p_{i+1}+p_{j+1})} \nonumber \\
&-& \frac{c_{1}(\lambda-\mu)}{b(\lambda-\mu)} B_{j}(\lambda) D_{ik}(\mu) (-1)^{p_{1}p_{i+1}+p_{j+1}(p_{1}+p_{i+1})}, \label{rule2} \\
B_{i}(\lambda) B_{j}(\mu) &=& \frac{1}{a_{1}(\lambda-\mu)} B_{l}(\mu) B_{q}(\lambda) {\check{r}^{(1)}(\lambda-\mu)}_{lq}^{ij} (-1)^{p_{1}(p_{i+1}+p_{l+1})+p_{1}},\label{rule3}
\ear
\end{subequations}
where ${\check{r}^{(1)}}(\lambda)_{\alpha\beta}^{\gamma\delta}$ are the elements of the $U_{q}[Sl(n-1+p_{1}|m-p_{1})]$ $R$-matrix (\ref{rmatr}). Here we remark that such elements are defined by  $\displaystyle \check{r}_{ab}^{(1)}(\lambda)=\sum_{\alpha\beta\gamma\delta=1}^{n+m-1} \check{r}^{(1)}(\lambda)_{\alpha\beta}^{\gamma\delta} \hat{e}_{\alpha \gamma}^{(a)}\otimes \hat{e}_{\beta \delta}^{(b)}$ and the new parities $p_{\alpha}^{(1)}$ defining the auxiliary matrix $\check{r}_{ab}^{(1)}(\lambda)$ are therefore $p_{\alpha}^{(1)}=p_{\alpha+1}$ for $\alpha=1,\dots, n+m-1$.

By iterating the fields $A(\lambda)$ and $D_{ii}(\lambda)$ over the multiparticle state (\ref{ansatz}) we generate states proportional to $\ket{\phi}$ as well as those that are not usually denominated unwanted terms. The former terms 
contribute to the eigenvalue $\Lambda(\lambda)$ and are obtained by keeping only the first term of the commutation rules (\ref{rule1}) and (\ref{rule2}). The unwanted terms, however, appear as a consequence of the second part of the relations (\ref{rule1}) and (\ref{rule2}) when the variables $\lambda_{k}^{(1)}$ are exchanged with the spectral parameter $\lambda$, which can be collected in closed form thanks to Eq.(\ref{rule3}). Since this procedure is by now standard in the literature we just list the main results. The corresponding eigenvalues are given by
\EQ
\Lambda(\lambda) =(-1)^{p_{1}}
\left[ f_{1}(\lambda) \right]^{L_{1}}\left[ g_{1}(\lambda) \right]^{L_{2}}
\prod_{j=1}^{\bar{m}_{1}}  \frac{a_{1}(\lambda^{(1)}_{j}-\lambda)}{b(\lambda^{(1)}_{j}-\lambda)}
+ \prod_{j=1}^{\bar{m}_{1}}
\frac{1}{b(\lambda-\lambda^{(1)}_{j})}
\Lambda^{(1)}(\lambda,\{\lambda^{(1)}_{k}\}),
\label{gamaalmost}
\EN
provided that the rapidities $\{\lambda_{k}^{(1)}\}$ satisfy the following Bethe ansatz equations
\bear
&&\left[f_{1}(\lambda^{(1)}_{i}) \right]^{L_{1}} \left[ g_{1}(\lambda^{(1)}_{i}) \right]^{L_{2}}
\prod_{\stackrel{j=1}{j \neq i}}^{\bar{m}_{1}} b(\lambda^{(1)}_{i}-\lambda^{(1)}_{j})
\frac{a_{1}(\lambda^{(1)}_{j}-\lambda^{(1)}_{i})}{b(\lambda^{(1)}_{j}-\lambda^{(1)}_{i})}  =
\Lambda^{(1)}(\lambda = \lambda^{(1)}_{i},\{\lambda^{(1)}_{k}\}), \nonumber \\
\ear
for $i=1, \dots, \bar{m}_{1}$.

The auxiliary function $\Lambda^{(1)}(\lambda,\{\lambda_{k}^{(1)} \})$ turns out to be the eigenvalues of an inhomogeneous transfer matrix, namely
\EQ
T^{(1)}(\lambda,\{\lambda^{(1)}_{k}\})^{\alpha_{1} \dots \alpha_{\bar{m}_{1}}}_{\beta_{1} \dots \beta_{\bar{m}_{1}}} {\cal F}^{\alpha_{\bar{m}_{1}} \dots \alpha_{1}}\ket{0} =
\Lambda^{(1)}(\lambda,\{\lambda^{(1)}_{k}\}) {\cal F}^{\beta_{\bar{m}_{1}} \dots \beta_{1}} \ket{0},
\label{NEST}
\EN
whose matrix elements are defined as
\bear
 T^{(1)}(\lambda,\{ \lambda_{k}^{(1)} \})_{\beta_{1}\dots \beta_{\bar{m}_{1}}}^{\alpha_{1}\dots \alpha_{\bar{m}_{1}}} &=& (-1)^{p_{i+1}(1+\sum_{k=1}^{\bar{m}_{1}} p_{\beta_{k}+1}) + p_{1} \sum_{k=1}^{\bar{m}_{1}-1}
p_{\delta_{k}+1} + (\bar{m}_{1}-1) p_{1}p_{i+1} }
 {\check{r}^{(1)}}(\lambda-\lambda_{1}^{(1)})_{\beta_{1} \delta_{1}}^{i \alpha_{1}} \dots \nonumber\\ &&{\check{r}^{(1)}}(\lambda-\lambda_{\bar{m}_{1}-1}^{(1)})_{\beta_{\bar{m}_{1}-1} \delta_{\bar{m}_{1}-1}}^{\delta_{\bar{m}_{1}-2} \alpha_{\bar{m}_{1}-1}}
{\check{r}^{(1)}}(\lambda-\lambda_{\bar{m}_{1}}^{(1)})_{\beta_{\bar{m}_{1}} \delta_{\bar{m}_{1}}}^{\delta_{\bar{m}_{1}-1} \alpha_{\bar{m}_{1}}} D_{i \delta_{\bar{m}_{1}}}(\lambda).
\label{auxtransf}
\ear

This completes the first step of the Bethe ansatz analysis since we still need to determine the eigenvalues $\Lambda^{(1)}(\lambda,\{\lambda_{k}^{(1)}\})$. In the next section we will discuss how this problem can be solved in terms of recurrence relations.

\section{Eigenvalues and Bethe ansatz equations}\label{eingBA}

Here we are concerned with the diagonalization of the auxiliary eigenvalue problem (\ref{NEST}) whose solution needs a second algebraic Bethe ansatz analysis. The transfer matrix (\ref{auxtransf}) can be rewritten as $T^{(1)}(\lambda,\{\lambda_{k}^{(1)} \})=\str_{{\cal A}^{(1)}}{ \left[ {\cal T}_{{\cal A}^{(1)}}^{(1)}(\lambda,\{\lambda_{k}^{(1)} \})  \right]}$ with ${\cal A}^{(1)} \equiv C^{n+m-1}$ such that its  correspondent monodromy matrix is given by
\EQ
{\cal T}_{{\cal A}^{(1)}}^{(1)}(\lambda,\{\lambda_{k}^{(1)} \})= \vec{D}_{{\cal A}^{(1)}}(\lambda) . r_{{\cal A}^{(1)}\bar{m}_{1}}^{(1)}(\lambda-\lambda_{\bar{m}_{1}}^{(1)}) r_{{\cal A}^{(1)}\bar{m}_{1}-1}^{(1)}(\lambda-\lambda_{\bar{m}_{1}-1}^{(1)}) \dots r_{{\cal A}^{(1)}1}^{(1)}(\lambda-\lambda_{1}^{(1)}),
\label{mono1}
\EN
where $r_{{\cal A}^{(1)}j}^{(1)}(\lambda)=P_{{\cal A}^{(1)}j} \check{r}_{{\cal A}^{(1)}j}^{(1)}(\lambda)$ and $\vec{D}_{{\cal A}^{(1)}}(\lambda)$ is a $(n+m-1)\times(n+m-1)$ matrix whose elements are the operators $D_{ij}(\lambda)$.  We also recall the products in (\ref{mono1}) are taken in the auxiliary space ${\cal A}^{(1)}$.

An important property of this auxiliary monodromy operator is that it satisfies the following intertwining relation
\EQ
\check{r}^{(1)}(\lambda-\mu) {\cal T}_{{\cal A}^{(1)}}^{(1)}(\lambda,\{\lambda_{k}^{(1)}\}) \stackrel{s_{1}}{\otimes} {\cal T}_{{\cal
A}^{(1)}}^{(1)}(\mu,\{\lambda_{k}^{(1)}\}) = {\cal T}_{{\cal A}^{(1)}}^{(1)}(\mu,\{\lambda_{k}^{(1)}\}) \stackrel{s_{1}}{\otimes} {\cal T}_{{\cal A}^{(1)}}^{(1)}(\lambda,\{\lambda_{k}^{(1)}\}) \check{r}^{(1)}(\lambda-\mu),
\label{alg1}
\EN
where now $\stackrel{s_{1}}{\otimes}$ stands for the supertensor product with parities $p_{\alpha}^{(1)}$.

The next step is to exhibit a pseudovacuum state $\ket{0}^{(1)}$ on which the monodromy matrix ${\cal T}_{{\cal A}^{(1)}}^{(1)}(\lambda,\{ \lambda_{k}^{(1)} \})$ acts as an upper triangular matrix. One way to achieve this is by a suitable combination of vectors acting on the tensor  product space $\prod_{j=1}^{L_{1}+L_{2}} \otimes  C_{j}^{n+m} \otimes \prod_{j=1}^{\bar{m}_{1}}\otimes C_{j}^{n+m-1}$ \cite{AB}. Once the first part of this tensor product was already determined as being $\ket{0}$, the second one should be set up provided the triangularity property is reached. It turns out that such appropriate state is
\EQ
\ket{0}^{(1)}=\ket{0} \otimes \prod_{i=1}^{\bar{m}_{1}} \otimes \left( \begin{array}{c}
1 \\
0 \\
\vdots
\\
0
\end{array} \right)_{(n+m-1)},
\label{newstate}
\EN
where the monodromy matrix elements of ${\cal T}_{{\cal A}^{(1)}}^{(1)}(\lambda,\{\lambda_{k}^{(1)} \})$ proportional to $D_{ij}(\lambda)$ act on the first part of $\ket{0}^{(1)}$ while the remaining operators act on the second part of the tensor product (\ref{newstate}).

In fact, by writing ${\cal T}_{{\cal A}^{(1)}}^{(1)}(\lambda,\{\lambda_{k}^{(1)} \})$ in the auxiliary space ${\cal A}^{(1)}$ as
\EQ
{\cal T}_{{\cal A}^{(1)}}^{(1)}(\lambda,\{\lambda_{k}^{(1)} \})=\left(
\begin{array}{cccc}
A^{(1)}(\lambda) & B_{1}^{(1)}(\lambda) & \cdots & B_{n+m-2}^{(1)}(\lambda) \\
C_{1}^{(1)}(\lambda) & D_{1,1}^{(1)}(\lambda) & \cdots & D_{1, n+m-2}^{(1)}(\lambda) \\
\vdots & \vdots & \ddots & \cdots \\
C_{n+m-2}^{(1)}(\lambda) & D_{n+m-2, 1}^{(1)}(\lambda) & \cdots & D_{n+m-2, n+m-2}^{(1)}(\lambda)
\end{array}\right)_{(n+m-1)\times (n+m-1)},
\EN
one can derive the following relations
\bear
A^{(1)}(\lambda,\{\lambda_{k}^{(1)}\})\ket{0}^{(1)} &=&
\left[ f_{2}(\lambda)\right]^{L_{1}} \left[g_{2}(\lambda)\right]^{L_{2}} \prod_{j=1}^{\bar{m}_{1}}a_{2}(\lambda-\lambda_{j}^{(1)}) \ket{0}^{(1)},
\\
D^{(1)}_{ii}(\lambda,\{\lambda_{k}^{(1)}\})\ket{0}^{(1)} &=& \left[ f_{i+2}(\lambda)\right]^{L_{1}} \left[
g_{i+2}(\lambda)\right]^{L_{2}} \prod_{j=1}^{\bar{m}_{1}} b(\lambda-\lambda_{j}^{(1)}) \ket{0}^{(1)}, \nonumber\\
i=1,\dots,n+m-2
\ear
\bear
D^{(1)}_{i,j}(\lambda,\{\lambda_{k}^{(1)}\})\ket{0}^{(1)}
&=& 0, ~~~ i\neq j, j \neq n+m-2, \\
D^{(1)}_{i,n+m-2}(\lambda,\{\lambda_{k}^{(1)}\})\ket{0}^{(1)}
&=& \#, ~~~ i\neq n+m-2, \\
C^{(1)}_{i}(\lambda,\{\lambda_{k}^{(1)}\})\ket{0}^{(1)} &=& 0 ~~~ i=1,n+m-2.
\ear

From now on the basic steps in order to diagonalize the transfer matrix $T^{(1)}(\lambda,\{\lambda_{k}^{(1)} \})$ become similar to those described in the previous section. One introduces a second multiparticle state $\ket{\phi}^{(1)}=B_{\alpha_{1}}^{(1)}(\lambda_{1}^{(2)},\{\lambda_{k}^{(1)} \}) \dots B_{\alpha_{\bar{m}_{2}}}^{(1)}(\lambda_{\bar{m}_{2}}^{(2)},\{\lambda_{k}^{(1)} \}) {\cal F}_{(1)}^{\alpha_{\bar{m}_{2}}\dots \alpha_{\bar{m}_{1}}} \ket{0}^{(1)}$ parameterized by a new set of inhomogeneities $\lambda_{1}^{(2)}\dots \lambda_{\bar{m}_{2}}^{(2)}$. Thanks to the Yang-Baxter algebra (\ref{alg1}) satisfied by ${\cal T}_{{\cal A}^{(1)}}^{(1)}(\lambda,\{\lambda_{k}^{(1)} \})$ the structure of the commutation rules between the new diagonal fields $A^{(1)}(\lambda,\{ \lambda_{k}^{(1)} \})$ and $D_{ii}^{(1)}(\lambda,\{ \lambda_{k}^{(1)} \})$ are exactly the same as that presented in section \ref{ABA}. At a general level $l$ each operator $\hat{O}(\lambda)$ in Eqs.(\ref{rule1}-\ref{rule3}) is substituted by its corresponding $\hat{O}^{(l)}(\lambda,\{\lambda_{k}^{(l)} \})$. The auxiliary $R$-matrix 
$\check{r}_{ab}^{(l)}(\lambda)$ is that of the $\displaystyle U_{q}\left[Sl(n-l+\sum_{k=1}^{l}p_{k}|m-\sum_{k=1}^{l}p_{k})\right]$ vertex model and the parities $p_{\alpha}^{(l)}=p_{\alpha+1}^{(l-1)}$ for $\alpha=1,\dots, n+m-l$. As a consequence of that the eigenvalues at subsequent steps $l$ and $l+1$ are going to satisfy a recurrence relation similar to that of Eq.(\ref{gamaalmost}). More precisely, for a given $l\geq 1$ we find that
\bear
\Lambda^{(l)}(\lambda,\{\lambda_{k}^{(l)} \}) =(-1)^{p_{l+1}}
\left[ f_{l+1}(\lambda) \right]^{L_{1}} \left[ g_{l+1}(\lambda)
\right]^{L_{2}}
\prod_{j=1}^{\bar{m}_{l}} a_{l+1}(\lambda-\lambda_{j}^{(l)}) \prod_{j=1}^{\bar{m}_{l+1}}  \frac{a_{l+1}(\lambda^{(l+1)}_{j}-\lambda)}{b(\lambda^{(l+1)}_{j}-\lambda)} \nonumber \\
 +
 \prod_{j=1}^{\bar{m}_{l}} b(\lambda-\lambda_{j}^{(l)}) \prod_{j=1}^{\bar{m}_{l+1}}
\frac{1}{b(\lambda-\lambda^{(l+1)}_{j})}
\Lambda^{(l+1)}(\lambda,\{\lambda^{(1)}_{k},\dots,\lambda^{(l)}_{k}\}),
\label{einrecl}
\ear
provided that the set of variables $\{\lambda_{k}^{(l+1)} \}$ that parameterize the eigenvectors of the inhomogeneous transfer matrix $T^{(l)}(\lambda,\{\lambda_{1}^{(l)}, \dots, \lambda_{\bar{m}_{l}}^{(l)} \})$ satisfy the following Bethe ansatz equations
\bear
&&\left[
f_{l+1}(\lambda_{i}^{(l+1)}) \right]^{L_{1}} \left[
g_{l+1}(\lambda_{i}^{(l+1)}) \right]^{L_{2}} \prod_{j=1}^{\bar{m}_{l}}
\frac{a_{l+1}(\lambda^{(l+1)}_{i}-\lambda^{(l)}_{j})}{b(\lambda^{(l+1)}_{i}-\lambda^{(l)}_{j})}
\prod_{\stackrel{j=1}{j \neq i}}^{\bar{m}_{l+1}}
b(\lambda^{(l+1)}_{i}-\lambda^{(l+1)}_{j})
\frac{a_{l+1}(\lambda^{(l+1)}_{j}-\lambda^{(l+1)}_{i})}{b(\lambda^{(l+1)}_{j}-\lambda^{(l+1)}_{i})}  = \nonumber \\
&&\Lambda^{(l+1)}(\lambda =
\lambda^{(l+1)}_{i},\{\lambda^{(1)}_{k},\dots,
\lambda^{(l+1)}_{k}\}), ~~~ i=1, \dots, \bar{m}_{l+1},
\label{BArecl}
\ear
for $l=1,\dots,n+m-3$.

Those equations can be iterated starting with $l=1$ until we reach the nested Bethe ansatz level $l=n+m-3$. In the last step $l=n+m-2$ one has to deal with a diagonalization of six vertex like model. In this case the final result for the eigenvalue $\Lambda^{(n+m-2)}(\lambda,\{\lambda_{k}^{(n+m-2)} \})$ is
\EQ
\Lambda^{(n+m-2)}(\lambda,\{\lambda_{k}^{(n+m-2)} \}) =
\EN
\bear
(-1)^{p_{n+m-1}} \left[ f_{n+m-1}(\lambda) \right]^{L_{1}} \left[ g_{n+m-1}(\lambda) \right]^{L_{2}}
\prod_{j=1}^{\bar{m}_{n+m-2}} a_{n+m-1}(\lambda-\lambda_{j}^{(n+m-2)}) \prod_{j=1}^{\bar{m}_{n+m-1}}  \frac{a_{n+m-1}(\lambda^{(n+m-1)}_{j}-\lambda)}{b(\lambda^{(n+m-1)}_{j}-\lambda)} \nonumber \\
+(-1)^{p_{n+m}} \left[ f_{n+m}(\lambda) \right]^{L_{1}} \left[ g_{n+m}(\lambda) \right]^{L_{2}} \prod_{j=1}^{\bar{m}_{n+m-2}} b(\lambda-\lambda_{j}^{(n+m-2)}) \prod_{j=1}^{\bar{m}_{n+m-1}}\frac{a_{n+m}(\lambda-\lambda^{(n+m-1)}_{j})}{b(\lambda-\lambda^{(n+m-1)}_{j})}, \nonumber
\label{ein6ver}
\ear
whose Bethe ansatz equations for the last variables $\{\lambda_{k}^{(n+m-1)} \}$ are
\bear
&&\left[ \frac{f_{n+m-1}(\lambda_{i}^{(n+m-1)})}{f_{n+m}(\lambda_{i}^{(n+m-1)})} \right]^{L_{1}} \left[ \frac{g_{n+m-1}(\lambda_{i}^{(n+m-1)})}{g_{n+m}(\lambda_{i}^{(n+m-1)})} \right]^{L_{2}} = \nonumber \\
&&\prod_{j=1}^{\bar{m}_{n+m-2}} \frac{b(\lambda^{(n+m-1)}_{i}-\lambda^{(n+m-2)}_{j})}{a_{n+m-1}(\lambda^{(n+m-1)}_{i}-\lambda^{(n+m-2)}_{j})}
\prod_{\stackrel{j=1}{j \neq i}}^{\bar{m}_{l+1}} \frac{a_{n+m}(\lambda^{(n+m-1)}_{i}-\lambda^{(n+m-1)}_{j})}{a_{n+m-1}(\lambda^{(n+m-1)}_{j}-\lambda^{(n+m-1)}_{i})}  \frac{b(\lambda^{(n+m-1)}_{j}-\lambda^{(n+m-1)}_{i})}{b(\lambda^{(n+m-1)}_{i}-\lambda^{(n+m-1)}_{j})}. \nonumber \\
\label{BA6ver}
\ear

We have now reached a point in which the results of section \ref{vertex} and \ref{ABA} can be put altogether. Combining  Eq.(\ref{gamaalmost}) with the iteration of Eq.(\ref{einrecl}) until $l=n+m-3$ and finally with the help of the expression (\ref{ein6ver}) we find that the eigenvalues $\Lambda(\lambda,\{\lambda_{k}^{(l)} \})$ are
\bear
&&\Lambda(\lambda;\{\lambda_{i}^{(1)}\}, \dots, \{\lambda_{i}^{(n+m-1)}\})=
(-1)^{p_{1}}\left[f_{1}(\lambda)\right]^{L_{1}}  \left[g_{1}(\lambda) \right]^{L_{2}} \prod_{j=1}^{\bar{m}_{1}} \frac{a_{1}(\lambda_{j}^{(1)}-\lambda)}{b(\lambda_{j}^{(1)}-\lambda)} \nonumber \\
&& + \sum_{k=2}^{n+m-1} (-1)^{p_{k}} \left[f_{k}(\lambda)\right]^{L_{1}} \left[g_{k}(\lambda)\right]^{L_{2}} \prod_{j=1}^{\bar{m}_{k-1}} \frac{a_{k}(\lambda-\lambda_{j}^{(k-1)})}{b(\lambda-\lambda_{j}^{(k-1)})}
\prod_{j=1}^{\bar{m}_{k}} \frac{a_{k}(\lambda_{j}^{(k)}-\lambda)}{b(\lambda_{j}^{(k)}-\lambda)}
\label{presque} \\
&& +(-1)^{p_{n+m}}\left[ f_{n+m}(\lambda)\right]^{L_{1}} \left[ g_{n+m}(\lambda)\right]^{L_{2}}
\prod_{j=1}^{ \bar{m}_{n+m-1} } \frac{a_{n+m}(\lambda-\lambda_{j}^{(n+m-1)})}{b(\lambda-\lambda_{j}^{(n+m-1)})},
 \nonumber
\ear
while the corresponding Bethe ansatz equations become
\bear
\left[\frac{f_{1}(\lambda_{i}^{(1)})}{f_{2}(\lambda_{i}^{(1)})}\right]^{L_{1}} \left[\frac{g_{1}(\lambda_{i}^{(1)})}{g_{2}(\lambda_{i}^{(1)})}\right]^{L_{2}} =
\prod_{\stackrel{j=1}{j \neq i}}^{\bar{m}_{1}}  \frac{a_{2}(\lambda_{i}^{(1)}-\lambda_{j}^{(1)})}{a_{1}(\lambda_{j}^{(1)}-\lambda_{i}^{(1)})} \frac{b(\lambda_{j}^{(1)}-\lambda_{i}^{(1)})}{b(\lambda_{i}^{(1)}-\lambda_{j}^{(1)})}
\prod_{j=1}^{\bar{m}_{2}} \frac{a_{2}(\lambda_{j}^{(2)}-\lambda_{i}^{(1)})}{b(\lambda_{j}^{(2)}-\lambda_{i}^{(1)})},
\label{firsteq} \\
\ear
\bear
&&\left[\frac{f_{k}(\lambda_{i}^{(k)})}{f_{k+1}(\lambda_{i}^{(k)})}\right]^{L_{1}} \left[\frac{g_{k}(\lambda_{i}^{(k)})}{g_{k+1}(\lambda_{i}^{(k)})}\right]^{L_{2}} = \nonumber \\
&&\prod_{j=1}^{\bar{m}_{k-1}} \frac{ b(\lambda_{i}^{(k)}-\lambda_{j}^{(k-1)})}{ a_{k}(\lambda_{i}^{(k)}-\lambda_{j}^{(k-1)})} \prod_{\stackrel{j=1}{j \neq i}}^{\bar{m}_{k}} \frac{a_{k+1}(\lambda_{i}^{(k)}-\lambda_{j}^{(k)})}{a_{k}(\lambda_{j}^{(k)}-\lambda_{i}^{(k)})}
\frac{b(\lambda_{j}^{(k)}-\lambda_{i}^{(k)})}{b(\lambda_{i}^{(k)}-\lambda_{j}^{(k)})}
\prod_{j=1}^{\bar{m}_{k+1}}
\frac{a_{k+1}(\lambda_{j}^{(k+1)}-\lambda_{i}^{(k)})}{b(\lambda_{j}^{(k+1)}-\lambda_{i}^{(k)})},
\ear
for $k=2, \dots, n+m-2$
\bear
&&\left[ \frac{f_{n+m-1}(\lambda_{i}^{(n+m-1)})}{f_{n+m}(\lambda_{i}^{(n+m-1)})} \right]^{L_{1}}
\left[ \frac{g_{n+m-1}(\lambda_{i}^{(n+m-1)})}{g_{n+m}(\lambda_{i}^{(n+m-1)})} \right]^{L_{2}}
= \prod_{j=1}^{\bar{m}_{n+m-2}} \frac{b(\lambda_{i}^{(n+m-1)}-\lambda_{j}^{(n+m-2)})}{a_{n+m-1}(\lambda_{i}^{(n+m-1)}-\lambda_{j}^{(n+m-2)})} \nonumber \\
&& \times\prod_{\stackrel{j=1}{j \neq i}}^{\bar{m}_{n+m-1}} \frac{a_{n+m}(\lambda_{i}^{(n+m-1)}-\lambda_{j}^{(n+m-1)})}{a_{n+m-1}(\lambda_{j}^{(n+m-1)}-\lambda_{i}^{(n+m-1)})}
\frac{b(\lambda_{j}^{(n+m-1)}-\lambda_{i}^{(n+m-1)})}{b(\lambda_{i}^{(n+m-1)}-\lambda_{j}^{(n+m-1)})}.
\label{lasteq}
\ear

Up to this point our analysis has been carried out for arbitrary dressing 
functions $f_{i}(\lambda)$ and $g_{i}(\lambda)$. In fact, the main results (\ref{presque}-\ref{lasteq}) 
are expected to be valid for any representations such that reference states with the 
properties (\ref{triang1}-\ref{triang2}) could be exhibited. We note that our results for the Bethe ansatz
equations (\ref{firsteq}-\ref{lasteq}) are equivalent to the analyticity  
of the eigenvalues (\ref{presque}) as function
of the rapidities  $\{\lambda_j^{(1)}\},\cdots,
\{\lambda_j^{(n+m-1)}\}$. The  structure 
of Eqs.(\ref{firsteq}-\ref{lasteq}) turns out also to be in accordance 
with that generally postulated in the realm of the analytical
Bethe ansatz  approach \cite{ANAL} 
for the $Sl(n|m)$ symmetry and its quantum affine extension  \cite{TEZ,MOR}. The explicit  construction
of the transfer matrix eigenvectors as presented here is, however, beyond the scope of the analytical
Bethe ansatz.

We now turn our attention to the specific case 
of the vector and dual representations of $U_{q}\left[Sl(n|m) \right]$. In this situation  
further simplifications can still be 
implemented in Eqs.(\ref{presque}-\ref{lasteq}) to bring them to assume a more traditional symmetrical form. This is 
implemented by performing the shift $\lambda_{j}^{(l)} \rightarrow \lambda_{j}^{(l)}- \delta^{(l)}$ where $\delta^{(l)}=i\frac{\gamma}{2}\sum_{k=1}^{l}(-1)^{p_{k}}$ and $q=e^{-i\gamma}$. Substituting the explicit expressions for the weights in Eqs.(\ref{presque}-\ref{lasteq}) and after some long but straightforward simplifications we find that
\bear
&&\Lambda(\lambda;\{\lambda_{i}^{(1)}\}, \dots, \{\lambda_{i}^{(n+m-1)}\})=
(-1)^{p_{1}}\left[a_{1}(\lambda)\right]^{L_{1}}  \left[\bar{b}(\lambda+\rho) \right]^{L_{2}} \prod_{j=1}^{\bar{m}_{1}} \frac{\sinh{\left[ \lambda_{j}^{(1)}-\lambda-\delta^{(1)} +i\gamma (-1)^{p_{1}}\right]}}{\sinh{\left[\lambda_{j}^{(1)}-\lambda-\delta^{(1)}\right]}} \nonumber \\
&& + \sum_{k=2}^{n+m-1} (-1)^{p_{k}} \left[b(\lambda)\right]^{L_{1}} \left[\bar{b}(\lambda+\rho)\right]^{L_{2}} \nonumber \\
&&\times \prod_{j=1}^{\bar{m}_{k-1}} \frac{\sinh{\left[ \lambda-\lambda_{j}^{(k-1)}+\delta^{(k-1)} +i\gamma (-1)^{p_{k}}\right]}}{\sinh{\left[\lambda-\lambda_{j}^{(k-1)}+\delta^{(k-1)} \right]}}
\prod_{j=1}^{\bar{m}_{k}} \frac{\sinh{\left[(\lambda_{j}^{(k)}-\lambda-\delta^{(k)} +i\gamma (-1)^{p_{k}}\right]}}{\sinh{\left[\lambda_{j}^{(k)}-\lambda-\delta^{(k)}\right]}}
 \\
&& +(-1)^{p_{n+m}}\left[ b(\lambda)\right]^{L_{1}} \left[ \bar{a}_{n+m}(\lambda+\rho)\right]^{L_{2}}
\prod_{j=1}^{ \bar{m}_{n+m-1} } \frac{\sinh{\left[\lambda-\lambda_{j}^{(n+m-1)}+\delta^{(n+m-1)} +\gamma (-1)^{p_{n+m}} \right]}}{\sinh{\left[\lambda-\lambda_{j}^{(n+m-1)}+\delta^{(n+m-1)}\right]}},
 \nonumber
\ear
and now the non-linear equations for the variables $\{\lambda_{k}^{(l)} \}$ are
\EQ
\left\{\frac{\sinh{\left[\lambda_{i}^{(1)}+i\frac{\gamma}{2}(-1)^{p_{1}}\right]}}{\sinh{\left[\lambda_{i}^{(1)}-i\frac{\gamma}{2}(-1)^{p_{1}}\right]}}\right\}^{L_{1}}=
\prod_{\stackrel{j=1}{j \neq i}}^{\bar{m}_{1}}  \frac{\sinh{\left[\lambda_{i}^{(1)}-\lambda_{j}^{(1)}+i\gamma (-1)^{p_{2}}\right]}}{\sinh{\left[\lambda_{i}^{(1)}-\lambda_{j}^{(1)}-i\gamma(-1)^{p_{1}}\right]}} \prod_{j=1}^{\bar{m}_{2}} \frac{\sinh{\left[ \lambda_{j}^{(2)}-\lambda_{i}^{(1)}+i\frac{\gamma}{2}(-1)^{p_{2}}\right]}}{\sinh{\left[\lambda_{j}^{(2)}-\lambda_{i}^{(1)}-i\frac{\gamma}{2}(-1)^{p_{2}}\right]}},
\label{ulteqmar}
\EN
\bear
&&\prod_{j=1}^{\bar{m}_{k-1}} \frac{ \sinh{ \left[ \lambda_{i}^{(k)}-\lambda_{j}^{(k-1)}+i\frac{\gamma}{2} (-1)^{p_{k}} \right]}}{ \sinh{\left[ \lambda_{i}^{(k)}-\lambda_{j}^{(k-1)} - i\frac{\gamma}{2} (-1)^{p_{k}}  \right]}}= \\
&&\prod_{\stackrel{j=1}{j \neq i}}^{\bar{m}_{k}} \frac{\sinh{ \left[ \lambda_{i}^{(k)}-\lambda_{j}^{(k)} + i\gamma (-1)^{p_{k+1}} \right]}}{\sinh{ \left[ \lambda_{i}^{(k)}-\lambda_{j}^{(k)} - i\gamma (-1)^{p_{k}} \right]}}
\prod_{j=1}^{\bar{m}_{k+1}}
\frac{\sinh{ \left[\lambda_{j}^{(k+1)}-\lambda_{i}^{(k)} +i\frac{\gamma}{2} (-1)^{p_{k+1}} \right]}}{\sinh{\left[\lambda_{j}^{(k+1)}-\lambda_{i}^{(k)} -i\frac{\gamma}{2} (-1)^{p_{k+1}} \right]}}, \nonumber \\
&&k=2, \dots, n+m-2 \nonumber
\ear
\bear
&&\left\{ \frac{\sinh{ \left[ \lambda_{i}^{(n+m-1)} +i \frac{\gamma}{2} (-1)^{p_{n+m}} +\rho-i \frac{\gamma}{2}(n-m-2) \right]}}{\sinh{ \left[ \lambda_{i}^{(n+m-1)} -i \frac{\gamma}{2} (-1)^{p_{n+m}} +\rho-i \frac{\gamma}{2}(n-m-2) \right]}} \right\}^{L_{2}}=  \nonumber \\
&&\prod_{j=1}^{\bar{m}_{n+m-2}} \frac{\sinh{\left[ \lambda_{i}^{(n+m-1)}-\lambda_{j}^{(n+m-2)} -i\frac{\gamma}{2} (-1)^{p_{n+m-1}}\right]}}{\sinh{\left[ \lambda_{i}^{(n+m-1)}-\lambda_{j}^{(n+m-2)} +i\frac{\gamma}{2} (-1)^{p_{n+m-1}}\right]}}  \label{ulteq} \\
&&\times \prod_{\stackrel{j=1}{j \neq i}}^{\bar{m}_{n+m-1}} \frac{\sinh{\left[ \lambda_{i}^{(n+m-1)}-\lambda_{j}^{(n+m-1)} +i\gamma (-1)^{p_{n+m}} \right]}}{\sinh{\left[ \lambda_{i}^{(n+m-1)}-\lambda_{j}^{(n+m-1)}-i\gamma (-1)^{p_{n+m-1}}\right]}}. \nonumber
\ear

Note that at the special value $\rho=i\gamma\left(\frac{n-m}{2}-1 \right)$ the Bethe ansatz equations (\ref{ulteqmar}-\ref{ulteq})  becomes a symmetrical system of non-linear equations. We close this section mentioning that the eigenvalues ${\cal E}^{(i)}$ of the spin chain ${\cal H}^{(i)}$ discussed in section \ref{vertex} are
\EQ
{\cal E}^{(i)}=\sum_{i=1}^{\bar{m}_{1}}\frac{2\sinh{\left[i\gamma(-1)^{p_{1}}\right]}}{\cosh{\left[2\lambda_{i}^{(1)}\right]}-\cosh{\left[i\gamma(-1)^{p_{1}}\right]}} +{\cal C}^{(i)},
\EN
where $\lambda_{k}^{(1)}$ satisfy the Bethe ansatz equations (\ref{ulteqmar}-\ref{ulteq}) and the additive constants ${\cal C}^{(i)}$ are
\bear
{\cal C}^{(1)}&=&(L-1)\left\{\coth{\left[i\gamma (-1)^{p_{1}}\right]}-\coth{\left[i\gamma\right]} \right\} + \left\{\coth{\left[\rho+i\gamma\right]}-\coth{\left[\rho\right]}\right\}, \\
{\cal C}^{(2)}&=&\frac{L}{2}\left\{\coth{\left[i\gamma (-1)^{p_{1}}\right]}-\coth{\left[i\gamma\right]} \right\}  +
\frac{L}{2}\left\{\coth{\left[\rho+i\gamma\right]}-\coth{\left[\rho\right]}\right\}.
\ear

\section{Conclusions}\label{Conclusion}

In this paper we have studied a trigonometric integrable vertex model that combines the vector and dual representations of $U_{q}\left[Sl(n|m)\right]$. The transfer matrix eigenvalue problem was solved by means of the algebraic Bethe ansatz method for all $\frac{(n+m)!}{n!m!}$ possible gradings. From our results for the transfer matrix eigenvalues and Bethe ansatz equations one can in principle derive the thermodynamic free-energy \cite{TES}. The fact that we know the solution for the many possibilities of grading choices could be of great help, since we can choose the particular one whose Bethe ansatz roots topology are the less complicated as possible. In particular, it would be interesting to determine the classes of universality governing the criticality of the gapless regimes. Finding the dependence of the low-lying critical exponents with $n$, $m$ and $\gamma$ may be of utility to make connections with supersymmetric Wess-Zumino-Witten conformal field theory \cite{SA}. We hope to return to this problem in a future publication.

\section*{Acknowledgements}
The author G.A.P. Ribeiro thanks FAPESP (Funda\c c\~ao de Amparo \`a Pesquisa do Estado de S\~ao Paulo)
for financial support and L. Amico for the hospitality of the DMFCI, Universit\`a di Catania, Italy. The work of M.J. Martins has been support by the Brazilian Research Council-CNPq and FAPESP.

\addcontentsline{toc}{section}{Appendix A}
\section*{\bf Appendix A: Hamiltonian coefficients}
\setcounter{equation}{0}
\renewcommand{\theequation}{A.\arabic{equation}}

In this appendix we define the Hamiltonian coefficient in terms of the Boltzmann weights defined in section \ref{vertex}.

\EQ
\Theta_{\alpha,\beta}=
\begin{cases}
\left[\underbar{$a$}_{\alpha}(\rho)\dot{\bar{c}}_{1}(\rho) q^{-2 \delta_{\alpha,1}}  + \underbar{$c$}_{1}(\rho)\dot{\bar{a}}_{\beta}(\rho)  \zeta_{\alpha\beta} \right] (-1)^{p_{\beta}}, & \text{if } \alpha < \beta \\
\left[\underbar{$a$}_{\alpha}(\rho)\dot{\bar{c}}_{2}(\rho) q^{2(\delta_{\beta,1}-1)}  + \underbar{$c$}_{2}(\rho)\dot{\bar{a}}_{\beta}(\rho)  1/\zeta_{\beta\alpha} \right] (-1)^{p_{\beta}},  & \text{if } \alpha > \beta \\
\displaystyle\sum_{\gamma=1}^{n+m} \underbar{$c$}_{1}(\rho)\dot{\bar{c}}_{1}(\rho) (-1)^{p_{\beta}+p_{\gamma}} q^{-2\delta_{\gamma,1}} \zeta_{\alpha\gamma} &  \text{if } \alpha<\gamma; ~ \gamma<\beta \\
\displaystyle\sum_{\gamma=1}^{n+m} \underbar{$c$}_{1}(\rho)\dot{\bar{c}}_{2}(\rho) (-1)^{p_{\beta}+p_{\gamma}}  q^{2(\delta_{\beta,1}-1)} \zeta_{\alpha\gamma} &  \text{if } \alpha<\gamma; ~ \gamma > \beta \\
\displaystyle\sum_{\gamma=1}^{n+m} \underbar{$c$}_{2}(\rho)\dot{\bar{c}}_{1}(\rho) (-1)^{p_{\beta}+p_{\gamma}} q^{-2\delta_{\gamma,1}} 1/\zeta_{\gamma\alpha}  &  \text{if } \alpha >\gamma; \gamma<\beta \\
\displaystyle\sum_{\gamma=1}^{n+m} \underbar{$c$}_{2}(\rho)\dot{\bar{c}}_{2}(\rho) (-1)^{p_{\beta}+p_{\gamma}} q^{2(\delta_{\beta,1}-1)} 1/\zeta_{\gamma\alpha}  &  \text{if } \alpha>\gamma; ~ \gamma > \beta
\end{cases}
\EN

\bear
\Omega^{(1)}_{\alpha,\beta}&=&
\begin{cases}
\underbar{$a$}_{\alpha}(\rho)\dot{c}_{1}(0)\bar{a}_{\alpha}(\rho) & \text{if } \alpha > \beta \\
\underbar{$a$}_{\alpha}(\rho)\dot{c}_{2}(0)\bar{a}_{\alpha}(\rho) & \text{if } \alpha < \beta
\end{cases} \\
\Omega^{(2)}_{\alpha,\beta}&=&
\begin{cases}
\underbar{$a$}_{\alpha}(\rho)\dot{a}_{\alpha}(0)\bar{c}_{1}(\rho) (-1)^{p_{\alpha}+p_{\beta}} q^{-2\delta_{\alpha,1}}  & \text{if } \alpha < \beta \\
\underbar{$a$}_{\alpha}(\rho)\dot{a}_{\alpha}(0)\bar{c}_{2}(\rho) (-1)^{p_{\alpha}+p_{\beta}} q^{2(\delta_{\beta,1}-1)}  & \text{if } \alpha > \beta \\
\end{cases} \\
\Omega^{(3)}_{\alpha,\beta}&=&
\begin{cases}
\underbar{$c$}_{1}(\rho)\dot{a}_{\beta}(0)\bar{a}_{\beta}(\rho) \zeta_{\alpha\beta}   & \text{if } \alpha < \beta \\
\underbar{$c$}_{2}(\rho)\dot{a}_{\beta}(0)\bar{a}_{\beta}(\rho) 1/\zeta_{\beta\alpha}  & \text{if } \alpha > \beta \\
\end{cases}
\ear

\EQ
\Xi_{\alpha,\beta,\gamma}=
\begin{cases}
\underbar{$a$}_{\alpha}(\rho)\dot{c}_{1}(0)\bar{c}_{1}(\rho) (-1)^{p_{\beta}} q^{-2\delta_{\alpha,1}}  & \text{if } \alpha < \beta; ~ \alpha < \gamma \\
\underbar{$a$}_{\alpha}(\rho)\dot{c}_{1}(0)\bar{c}_{2}(\rho) (-1)^{p_{\beta}} q^{2(\delta_{\beta,1}-1)}  & \text{if } \alpha > \beta; ~ \alpha < \gamma \\
\underbar{$a$}_{\alpha}(\rho)\dot{c}_{2}(0)\bar{c}_{1}(\rho) (-1)^{p_{\beta}} q^{-2\delta_{\alpha,1}}  & \text{if } \alpha < \beta; ~ \alpha > \gamma \\
\underbar{$a$}_{\alpha}(\rho)\dot{c}_{2}(0)\bar{c}_{2}(\rho) (-1)^{p_{\beta}} q^{2(\delta_{\beta,1}-1)}  & \text{if } \alpha > \beta; ~ \alpha > \gamma \\
\underbar{$c$}_{1}(\rho)\dot{a}_{\gamma}(0)\bar{c}_{1}(\rho) (-1)^{p_{\beta}} q^{-2\delta_{\gamma,1}} \zeta_{\alpha\gamma}   & \text{if } \alpha < \gamma; ~ \beta > \gamma \\
\underbar{$c$}_{2}(\rho)\dot{a}_{\gamma}(0)\bar{c}_{1}(\rho) (-1)^{p_{\beta}} q^{-2\delta_{\gamma,1}} 1/\zeta_{\gamma\alpha}   & \text{if } \alpha > \gamma; ~ \beta > \gamma \\
\underbar{$c$}_{1}(\rho)\dot{a}_{\gamma}(0)\bar{c}_{2}(\rho) (-1)^{p_{\beta}} q^{2(\delta_{\beta,1}-1)} \zeta_{\alpha\gamma}   & \text{if } \alpha < \gamma; ~ \beta < \gamma \\
\underbar{$c$}_{2}(\rho)\dot{a}_{\gamma}(0)\bar{c}_{2}(\rho) (-1)^{p_{\beta}} q^{2(\delta_{\beta,1}-1)} 1/\zeta_{\gamma\alpha}   & \text{if } \alpha > \gamma; ~ \beta < \gamma \\
\underbar{$c$}_{1}(\rho)\dot{c}_{2}(0)\bar{a}_{\beta}(\rho) (-1)^{p_{\beta}} \zeta_{\alpha\beta}   & \text{if } \alpha < \beta; ~ \beta > \gamma \\
\underbar{$c$}_{1}(\rho)\dot{c}_{1}(0)\bar{a}_{\beta}(\rho) (-1)^{p_{\beta}} \zeta_{\alpha\beta}   & \text{if } \alpha < \beta; ~ \beta < \gamma \\
\underbar{$c$}_{2}(\rho)\dot{c}_{2}(0)\bar{a}_{\beta}(\rho) (-1)^{p_{\beta}} 1/\zeta_{\beta\alpha}   & \text{if } \alpha > \beta; ~ \beta > \gamma \\
\underbar{$c$}_{2}(\rho)\dot{c}_{1}(0)\bar{a}_{\beta}(\rho) (-1)^{p_{\beta}} 1/\zeta_{\beta\alpha}   & \text{if } \alpha > \beta; ~ \beta < \gamma \\
\displaystyle \sum_{\nu=1}^{n+m} \underbar{$c$}_{1}(\rho)\dot{c}_{2}(0)\bar{c}_{1}(\rho) (-1)^{p_{\beta}+p_{\nu}} q^{-2\delta_{\nu,1}} \zeta_{\alpha\nu}  & \text{if } \alpha < \nu; ~ \beta > \nu; ~ \gamma < \nu \\
\displaystyle \sum_{\nu=1}^{n+m} \underbar{$c$}_{1}(\rho)\dot{c}_{2}(0)\bar{c}_{2}(\rho) (-1)^{p_{\beta}+p_{\nu}} q^{2(\delta_{\beta,1}-1)} \zeta_{\alpha\nu}  & \text{if } \alpha < \nu; ~ \beta < \nu; ~ \gamma < \nu \\
\displaystyle \sum_{\nu=1}^{n+m} \underbar{$c$}_{1}(\rho)\dot{c}_{1}(0)\bar{c}_{1}(\rho) (-1)^{p_{\beta}+p_{\nu}} q^{-2\delta_{\nu,1}} \zeta_{\alpha\nu}  & \text{if } \alpha < \nu; ~ \beta > \nu; ~ \gamma > \nu \\
\displaystyle \sum_{\nu=1}^{n+m} \underbar{$c$}_{1}(\rho)\dot{c}_{1}(0)\bar{c}_{2}(\rho) (-1)^{p_{\beta}+p_{\nu}} q^{2(\delta_{\beta,1}-1)} \zeta_{\alpha\nu}  & \text{if } \alpha < \nu; ~ \beta < \nu; ~ \gamma > \nu \\
\displaystyle \sum_{\nu=1}^{n+m} \underbar{$c$}_{2}(\rho)\dot{c}_{2}(0)\bar{c}_{1}(\rho) (-1)^{p_{\beta}+p_{\nu}} q^{-2\delta_{\nu,1}} 1/\zeta_{\nu\alpha}  & \text{if } \alpha > \nu; ~ \beta > \nu; ~ \gamma < \nu \\
\displaystyle \sum_{\nu=1}^{n+m} \underbar{$c$}_{2}(\rho)\dot{c}_{2}(0)\bar{c}_{2}(\rho) (-1)^{p_{\beta}+p_{\nu}} q^{2(\delta_{\beta,1}-1)} 1/\zeta_{\nu\alpha}  & \text{if } \alpha > \nu; ~ \beta < \nu; ~ \gamma < \nu \\
\displaystyle \sum_{\nu=1}^{n+m} \underbar{$c$}_{2}(\rho)\dot{c}_{1}(0)\bar{c}_{1}(\rho) (-1)^{p_{\beta}+p_{\nu}} q^{-2\delta_{\nu,1}} 1/\zeta_{\nu\alpha}  & \text{if } \alpha > \nu; ~ \beta > \nu; ~ \gamma > \nu \\
\displaystyle \sum_{\nu=1}^{n+m} \underbar{$c$}_{2}(\rho)\dot{c}_{1}(0)\bar{c}_{2}(\rho) (-1)^{p_{\beta}+p_{\nu}} q^{2(\delta_{\beta,1}-1)} 1/\zeta_{\nu\alpha}  & \text{if } \alpha > \nu; ~ \beta < \nu; ~ \gamma > \nu \\
\end{cases}
\EN

\bear
\Gamma^{(1)}_{\alpha,\beta,\gamma}&=&
\begin{cases}
\underbar{$b$}(\rho)\dot{c}_{1}(0)\bar{b}(\rho) & \text{if } \alpha < \beta; ~ \alpha \neq \gamma \\
\underbar{$b$}(\rho)\dot{c}_{2}(0)\bar{b}(\rho) & \text{if } \alpha > \beta; ~ \alpha \neq \gamma
\end{cases} \\
\Gamma^{(2)}_{\alpha,\beta,\gamma}&=&
\begin{cases}
\underbar{$a$}_{\alpha}(\rho)\dot{b}(0)\bar{b}(\rho) (-1)^{p_{\alpha}p_{\beta}} \delta_{\alpha,\gamma} & \text{if } \alpha \neq \beta \\
\underbar{$b$}(\rho)\dot{b}(0)\bar{a}_{\beta}(\rho) (-1)^{p_{\alpha}p_{\beta}} \delta_{\beta,\gamma} & \text{if } \alpha \neq \beta \\
\underbar{$b$}(\rho)\dot{b}(0)\bar{b}(\rho) (-1)^{p_{\alpha}p_{\beta}}  & \text{if } \alpha \neq \beta; ~ \alpha \neq \gamma; ~ \beta \neq \gamma
\end{cases} \\
\Gamma^{(3)}_{\alpha,\beta,\gamma}&=&
\begin{cases}
\underbar{$b$}(\rho)\dot{b}(0)\bar{c}_{1}(\rho) (-1)^{p_{\alpha}p_{\gamma}+p_{\beta}} q^{-2\delta_{\gamma,1}} & \text{if } \alpha \neq \gamma; ~ \beta > \gamma \\
\underbar{$b$}(\rho)\dot{b}(0)\bar{c}_{2}(\rho) (-1)^{p_{\alpha}p_{\gamma}+p_{\beta}} q^{2(\delta_{\beta,1}-1)} & \text{if } \alpha \neq \gamma; ~ \beta < \gamma
\end{cases} \\
\Gamma^{(4)}_{\alpha,\beta,\gamma}&=&
\begin{cases}
\underbar{$c$}_{1}(\rho)\dot{b}(0)\bar{b}(\rho) (-1)^{p_{\gamma}p_{\beta}+p_{\gamma}} \zeta_{\alpha\gamma} & \text{if } \alpha < \gamma; ~ \beta \neq \gamma \\
\underbar{$c$}_{2}(\rho)\dot{b}(0)\bar{b}(\rho) (-1)^{p_{\gamma}p_{\beta}+p_{\gamma}} 1/\zeta_{\gamma\alpha} & \text{if } \alpha > \gamma; ~ \beta \neq \gamma
\end{cases}
\ear

\end{document}